\documentclass[12pt]{article}

\begin{document}

\begin{center} {\Large \bf Dephasing of interacting tunneling systems}\\
Peter Nalbach$^a$\\
$^a$Department of Physics, Stanford University, Stanford, CA 94305, USA\\
\today
\end{center}

\abstract{
We investigate the phase coherence time of weakly interacting
tunneling systems (TSs). We show that all neighbors of a given TS form
together with the TS of interest an entangled cluster as long as the
linewidth of the excitation of the neighbor is smaller than its
interaction energy with the TS of interest. Thus, the relaxation of
all neighbors in the cluster contributes to the dephasing of the TS of
interest. This mechanism dominates the transversal decay of the TSs
and it explains recent two-pulse echo experiments in which the
exponential decay rate could not be explained within spectral
diffusion consistent with internal friction data. However, since the
proposed mechanism predicts only an exponential decay, the Gaussian
like decay at short times remains unexplained.}


\section{Introduction}

The low temperature properties of glasses are governed by tunneling
systems (TSs) and their behavior is well described by the
phenomenological Tunneling Model (TM) \cite{Phillips72,PWA72} 
down to temperatures of about 100 mK. 
Therein, the TSs are considered as independent two-level systems
(TLSs) with energy splitting
$\epsilon=\sqrt{\Delta_0^2+\Delta^2}$. The tunneling splitting
$\Delta_0$ and the asymmetry $\Delta$ are assumed to be broadly
distributed. The TSs interchange energy with phonons via their elastic
moment leading to a relaxation towards thermal equilibrium. At
low temperatures the one-phonon process dominates yielding the
relaxation rate
$\gamma_l=\gamma_0\Delta_0^2\epsilon\coth\left(\beta\epsilon/2\right)$
\cite{Jaeck72} and the phase coherence time $\tau_p=1/\gamma_t=2/\gamma_l$. 

For temperatures below about 100 mK the tunneling model fails in various
aspects which is commonly attributed to small interactions between the
tunneling systems. Although the assumption that with decreasing
temperature the importance of interactions increases seems natural,
the theoretical understanding of the influence of interactions between
tunneling systems to their dynamics is still unsatisfactory.
The importance of including interactions was firstly realized in
polarization echo experiments \cite{SD0} which are analogous to magnetic
spin echos. For example in so-called two pulse echos one is able to
investigate directly the phase coherence time of a subset of TSs with
fixed energy splitting.
Thereby the decay of the phase coherence is
partly due to the one-phonon process but the dominating process is due
to the interactions between the TSs. All TSs with energy splittings smaller
than the temperature undergo continuously thermal transitions which
are accompanied by fluctuating strain fields in their
environment. Since every TSs under investigation within an echo
experiment will have many such neighbors, this leads to a fluctuating
energy splitting for these TSs. This process know as
{\sl spectral diffusion}  destroys the phase coherence of the echo
amplitude \cite{KlauAnd68}.

Within spectral diffusion one distinguish between two limits
\cite{BlaHal77}. At short
times $t_{12}\ll\tau_{min}(T):=1/\gamma_l(\epsilon$=$\Delta_0$=$T)$
the decay of the two-pulse echo amplitude $A(t_{12})$ is expected to be
non-exponential, varying with the time $t_{12}$ between the two
pulses like $\propto\exp(-(2t_{12}/\tau_{p1})^2)$ with
$\tau_{p1}=\sqrt{h\tau_{min}/(2\delta\epsilon)}$. 
Thereby $\tau_{min}$ is the minimal relaxation time (governed by the
one-phonon process) of thermal TSs, which contribute the most to
spectral diffusion, and
$\delta\epsilon=(\Delta/\epsilon)J_{min}P_0k_BT$ is the spectral width
of the energy splitting $\epsilon$ of a TS with asymmetry $\Delta$ due
to the interaction with its neighbors. $J_{min}$ is the minimal
coupling between the TSs and $P_0$ is the prefactor of the TM-distribution
$P(\Delta_0,\Delta)=P_0/\Delta_0$. The expression $J_{min}P_0k_BT$ is
approximately the mean coupling between thermal TSs.
In the long time regime $t_{12}\gg\tau_{min}$ one finds an
exponential decay: $\propto\exp(-2t_{12}/\tau_{p2})$ with
$\tau_{p2}=h/(4\delta\epsilon)$. 
From the experimental point of view both time regimes where found
\cite{EnssSD96} but the extracted minimal relaxation time $\tau_{min}$
is by several orders of magnitude shorter than the values extracted
form internal friction experiments.

A key assumption in the theory of spectral diffusion is that the
interaction $J$ with a neighboring TS is smaller
than the linewidth of the excitation energy of that
neighbor. Accordingly a flip of that neighbor is possible without
energy transfer between the two TSs and they can still be treated as
isolated. Due to the broad distribution of interaction energies we
expect that many neighbors will not fulfill that condition. In that
case, where the linewidth is smaller than the 
coupling, the TSs are {\sl entangled} and we have to treat
them as a coupled cluster.

In the following we investigate the phase coherence time of a TS
within such a cluster. 
\section{Interacting tunneling systems}
Since the couplings are considered as weak, it is convenient to
switch directly to the diagonal basis of the uncoupled TLSs leading to
the Hamiltonian 
\marginpar{Hd}
\begin{eqnarray}\label{Hd} \hat{H} &=& -\sum_i \frac{\epsilon_{i}}{2}\sigma_x^{(i)}
\,-\sum_{i\not= j} \bar{u}_i \bar{u}_j
J_{ij}\sigma_x^{(i)}\sigma_x^{(j)} \nonumber \\ 
&& \,-\sum_{i\not= j} \left\{ u_i u_j J_{ij}\sigma_z^{(i)}\sigma_z^{(j)} \,+
\bar{u}_i u_j J_{ij}\sigma_x^{(i)}\sigma_z^{(j)} + u_i \bar{u}_j
J_{ij}\sigma_z^{(i)}\sigma_x^{(j)}  
\right\} \nonumber
\end{eqnarray}
with $u_i=\Delta_{0i}/\epsilon_i$, $\bar{u}_i=\Delta_{i}/\epsilon_i$
and $\epsilon_i=\sqrt{\Delta_{0i}^2+\Delta_i^2}$.
The main approximation of the present paper is to discuss the four
coupling terms separately; thus, we are left with three cases: the
$\sigma_x^{(i)}\sigma_x^{(j)}$-, the $\sigma_z^{(i)}\sigma_z^{(j)}$-
and the $\sigma_z^{(i)}\sigma_x^{(j)}$-coupling\footnote{The
$\sigma_x^{(i)}\sigma_z^{(j)}$-coupling term is 
equivalent to the $\sigma_z^{(i)}\sigma_x^{(j)}$-term.}. 
This separation neglects correlation effects between the
different coupling terms but it reveals first order effects like a
perturbative approach. 

We find that the term $\propto\sigma_z^{(i)}\sigma_x^{(j)}$ only yield
corrections to 
the dynamics of the 
order $O((J/\epsilon)^2)$, thus they are negligble for weak
couplings. The same holds true for the {\sl transversal} coupling term 
$\propto\sigma_z^{(i)}\sigma_z^{(j)}$ unless the TSs are resonant,
meaning that they have comparable energy splitting $\epsilon_i\simeq
\epsilon_j$; 
to be precise, the difference should be smaller than the coupling
between them: $|\epsilon_i-\epsilon_j|\le J_{ij}$. 
Since the probability of such pairs is $\le 10^{-3}$, this term is also
neglectable. Burin et al. \cite{Burin} found that resonant pairs lead
to a new relaxation mechanism at lowest temperature in glasses. But
this effect bases inherently on the combination of the
$\sigma_x^{(i)}\sigma_x^{(j)}$- and the
$\sigma_z^{(i)}\sigma_z^{(j)}$-term and, thus, it is 
beyond the scope of the present paper.  
We emphasize that for weak coupling the most important coupling term
is the $\propto\sigma_x^{(i)}\sigma_x^{(j)}$-term leading to the
Hamiltonian: 
$\hat{H}=-\sum_i \frac{\epsilon_i}{2}\sigma_x^{(i)}
-\sum_{i\not= j}\bar{u}_i\bar{u}_jJ_{ij}\sigma_x^{(i)}\sigma_x^{(j)}$.

In order to describe dynamics exhibiting relaxation and phase
decoherence, we couple additionally each TSs to phonons. 
This problem is investigated in detail in ref. \cite{Nal01b}.
The Hamiltonian is diagonal and accordingly the
eigenstates of the many-body problem are the product states of the
isolated TSs. Nevertheless, the interaction leads to entanglement
between the TSs.

For example, consider a TS with energy splittings $\epsilon_1$
coupled just to one neighbor with splitting $\epsilon_2$ and an
interaction energy $J_{12}$. If we couple with an field to the TS 1,
the excitation energy $\omega=\epsilon_1\pm 2J_{12}$ depends on the
present state of the second TS. The dephasing or transversal rate of
the TS 1 depends also on the state of the second TS \cite{Nal01b}. As long
as the coupling energy $J_{12}$ is bigger 
than the linewidth $\gamma_2$ of the second TS (given by the
transversal rate of the isolated second TS), the dephasing rate for
the first TS is the sum
$\Gamma_1=\gamma_1+\gamma_{2,\uparrow/\downarrow}$ with the one-phonon
rate $\gamma_1$ of the isolated TS 1 and the decay rate of the second
TS due to phonon emission 
$\gamma_{2\downarrow}=\gamma_0u_2^2\epsilon_2^3(1+n(\epsilon_2))$
or phonon absorption
$\Gamma_{2\uparrow}=\gamma_0u_2^2\epsilon_2^3n(\epsilon_2)$
with the Bose factor $n(\epsilon_2)$. Here and in the following we
approximate $\epsilon_i\pm J_{ij}\simeq\epsilon_i$.
If $\gamma_2>J_{12}$ the coupling is negligble and the dynamics of
the first TS is independent of the second one. 

The same holds true for many neighbors of a given TS $\alpha$ leading
to a dephasing rate
\begin{equation} \Gamma_t \,=\, \gamma_{\alpha} \,+ 2\sum_{i \in \atop \{i|J_{\alpha
i}>\gamma_{i}\} } \Big\{ \begin{array}{cc} \gamma_{i\uparrow}
& \mbox{for TLS $i$ in the ground state} \\ \gamma_{i\downarrow}
& \mbox{for TLS $i$ in the excited state} \end{array} \; .
\end{equation}
Thereby we distinguish between neighboring TSs
fulfilling $J_{\alpha i}>\gamma_{i}$, which contributes to the rate of
TS $\alpha$ and the TSs whose coupling to the TS $\alpha$ is to weak. This
second group of neighbors causes spectral diffusion. Assuming
homogeneously distributed TSs and a 
dipole like interaction between them, we expect a distribution of
interaction energies like $P(J)\propto 1/J^2$ with equal probability
for both signs. Since the above introduced conditions only involve the
absolute values of the coupling energies we can neglect the sign of
the coupling in the following. In order to gain simple analytical
expressions we separate the neighbors sharply between thermally active
and inactive ones with the further approximation
$\exp(-\epsilon_i/(k_BT))\simeq\Theta(k_BT-\epsilon_i)$. Using the
tunneling model distribution we get for the dephasing
\begin{equation} \langle e^{-\Gamma t}\rangle \,\simeq\, e^{-\gamma_{\alpha}
t \,-\Gamma_{ww} t} \qquad\mbox{with}\quad \Gamma_{ww} \,=\,
2\bar{u}_\alpha (P_0J_{min})\, (k_BT)\,
\ln\left(\frac{k_BT}{\Delta_{0min}}\right)
\end{equation}
with the minimal interaction energy $J_{min}$ and the minimal
tunneling element $\Delta_{0min}$. Thereby $\langle\cdot\rangle$
denotes the thermal averaging of the neighbors. 
$(P_0k_BT)$ is the number of thermally active
TSs and $J_{min}$ is approximately the mean interaction
between two TSs: $\bar{J}\simeq J_{min}\ln(J_{max}/J_{min})$. 

Thus the entanglement between the TSs mediated by weak couplings
results in a considerably fastened dephasing.

\section{Discussion and Summary}

In our approach we divide the neighbors of a given TS $\alpha$ into two
groups. The TSs, fulfilling $J_{i\alpha}>\gamma_i$, form a cluster
with the TS $\alpha$ leading to an enhanced decoherence rate for the
TS $\alpha$. The dynamics of the other neighbors with $J_{i\alpha}<\gamma_i$
still cause spectral diffusion. The dephasing rate due to the
entangled cluster is comparable with the exponential decay
of spectral diffusion in the long time limit: $\Gamma_{ww}\simeq
1/\tau_{p2}$. 
In typical echo experiments the decay in the long time regime should
also be faster than the one of the short time regime since 
$\tau_{min}(T)>\tau_{p2}$. Accordingly we expect the decay
$\exp(-\Gamma_{ww} t)$ due to the entanglement to dominate the decay
studied in two pulse echo experiments.

So far, the present theory describes fairly well the temperature
dependence as well as the absolute value of the exponential decay
measured in two-pulse polarization echos in B$_2$O$_3$ \cite{EnssSD00}
and in SiO$_2$ (Suprasil I) \cite{EnssSD96}. But within this theory we
can not account for the initial non-exponential decay found at lowest
temperatures. At least, the crossover depends no longer on the minimal
relaxation time $\tau_{min}(T)$ of the thermal TSs. Thus the
inconsistency with data of the internal friction is resolved. 

{\bf Acknowledgements}

This work has been supported by the Alexander-von-Humboldt foundation.

\end{document}